\documentclass[journal, doublecolumn, 10pt]{IEEEtran}
\usepackage{setspace}
\usepackage{graphicx}
\usepackage{amsmath,amssymb,bm,amsfonts,amsthm}
\usepackage{color}
\usepackage{algorithm,algorithmicx}
\usepackage[noend]{algpseudocode}
\usepackage{multirow}
\usepackage{cite}
\usepackage{comment}
\usepackage{subfig}
\usepackage{graphicx}

\newtheorem*{remark}{Remark}

\renewenvironment{IEEEbiography}[1]
  {\IEEEbiographynophoto{#1}}
  {\endIEEEbiographynophoto}

\begin{document}
\title{Towards Resilient Access Equality for 6G Serverless p-LEO Satellite Networks}
\author{Shih-Chun~Lin, Chia-Hung Lin, Liang C. Chu, and Shao-Yu Lien 
\IEEEcompsocitemizethanks{
\IEEEcompsocthanksitem Shih-Chun Lin and Chia-Hung Lin are with North Carolina State University. Liang C. Chu is with Lockheed Martin Space Systems Company, Lockheed Martin Corporation.
Shao-Yu Lien is with National Chung Cheng University.
\IEEEcompsocthanksitem This work was supported by Lockheed Martin Space Systems Company, Lockheed Martin Corporation.
}
}
\markboth{Submitted for publication in the IEEE Commun.}%
\maketitle

\IEEEcompsoctitleabstractindextext{%
\begin{abstract}
Low earth orbit (LEO) mega-constellations, integrating government space systems and commercial practices, have emerged as enabling technologies for the sixth generation (6G) networks due to their good merits of global coverage and ubiquitous services for military and civilian use cases. However, convergent LEO-based satellite networking infrastructures still lack leveraging the synergy of space and terrestrial systems. This paper, therefore, extends conventional serverless cloud platforms with serverless edge learning architectures for 6G proliferated LEO (p-LEO) satellite ecosystems and provides a new distributed training design from a networking perspective. The proposed design dynamically orchestrates communications and computation functionalities and resources among heterogeneous physical units to efficiently fulfill multi-agent deep reinforcement learning for service-level agreements. Innovative ecosystem enhancements, including ultrabroadband access, anti-jammed transmissions, resilient networking, and related open challenges, are also investigated for end-to-end connectivity, communications, and learning performance.
\end{abstract}
}
\maketitle
\IEEEdisplaynotcompsoctitleabstractindextext
\IEEEpeerreviewmaketitle

\section{Introduction}\label{sec1}
Emerging proliferated low earth orbit (p-LEO) constellations~\cite{22TWCJi,21JSACMou,21DASCXu,22WCNCRoper,22TNSMKak,Di19WC} with both government and commercial satellites have been regarded as the most promising remedy to provide global coverage and ubiquitous wireless services, bridging the ever-existent digital divide via their global footprints. These broadband connectivities to the Internet allow people to access information and essential services, such as governmental and health systems, ubiquitously without suffering geographical limitations. The need for resilient access equality via LEO satellites' high communications capacity with ultra-wide ranges~\cite{Di19WC,22TNSMKak} is more urgent than ever, as we have abruptly switched to remote living in past years due to the COVID-19 pandemic. Meanwhile, distributed machine learning (ML) brought attentions to decentralized data sources. Such learning technology will likely address multi-dimensional resource allocations for integrated mega-constellations and the sixth generation (6G) networks~\cite{19TVTQiu,21JSACMou,Cao21TC}. However, the recent federated or collaborative satellites mainly work on their feasibility via algorithm implementations (e.g.,~\cite{21DASCXu,22WCNCRoper}). There is little investigation into the redundancy and tradeoff between computations and communications and the dedicated resource orchestrations to realize timely edge learners with efficient data processing. Also, few solutions exist to comprehensively evaluate distributed training performance concerning non-terrestrial networks' peculiarities, e.g., satellite access and multi-tier connected infrastructure. Architectural, management and operational changes are required to realize the ecosystems.

This article presents a serverless software-defined networking (SDN) architecture that dynamically orchestrates communications and computation resources for a diverse set of 6G service-level agreements (SLAs). It provides a multi-tier ML framework that uses a unified control platform to optimize networking and resource configurations according to space and ground tiers' interactions. This coherent framework, coupled with underlying software-defined infrastructure, focuses on practical constraints and peculiarities in each tier, such as heterogeneous computing capabilities of ground terminals and size, weight, and power (SWaP)-limited satellites, frequently-handover satellite access, and coverage-limited ground tier. Remarkably, the intelligence within multiple ML-SDN control engines (e.g., ground terminals and satellites with computing capabilities) can realize efficient broadband access for ground users concerning software-reconfigurable LEO satellites, data-driven approaches for unknown environments, and different decision timescales of each unit. The design can use multi-tier ML models to establish high-throughput, reliable end-to-end transmissions for global connectivity. It is noteworthy that our designs tightly align with the latest industry specifications. For instance, 3GPP Release 17 considers satellite mobility at different orbital heights to support non-terrestrial networks with 3GPP NR on the ground. Release 18 in 2022 creates 5G Advanced to include new intelligent, ML-enabled solutions to boost mobile broadband and verticals performance.

The proposed architecture and solutions will bring several benefits to 6G p-LEO satellite ecosystems, as follows:
\begin{enumerate}
\item We provide \textbf{a serverless SDN edge infrastructure}, where the above applications only need to care about function implementation without managing any underlying resources. Compared with existing ground hardware-based infrastructures that significantly delay innovation deployments, the proposed infrastructure introduces a serverless computing layer based on an abstraction of the ground-space ecosystem. Both application-defined interfaces and network resource management tools are developed to manage applications on-demand through dynamically instantiated containers and effectively utilize computing resources from distributed satellites and ground terminals.

\item We improve \textbf{the end-to-end learning performance} by dynamically adjusting workloads among ML-SDN control engines. For example, ground terminals with powerful computing capabilities can assist satellites' resource allocation tasks. The satellite learning model can also be transferred to its successor, preventing always training from scratch for SWaP-limited devices.

\item We enable \textbf{data-driven multi-user access control} for the ground-space eco-network. Serverless computing architectures provide infrastructure controllability to the multi-tier ML framework. This framework can establish efficient ultra-broadband sensing and communications between the two tiers to satisfy 6G requirements.

\item We achieve \textbf{reliable software-defined internetworking} through spectrum harmonization and hyper-connectivity. New networking designs are realized to address heterogeneity, scalability, performance, and reliability fully. For example, we investigate (i) anti-jammed multiple-input-multiple-output (MIMO) p-LEOs with Ka-bands, MIMO beamforming, and frequency-hopping techniques, (ii) ``Space Highway" to leverage mega-constellation structure and enhance global transmission capacities, and (iii) cyber-harden opportunistic multipath routing to solve long delay and cyber-attacks.
\end{enumerate}
Therefore, our innovations significantly enhance end-to-end performance and impact future human society in isolated or remote communities and landlocked areas with limited infrastructure investments. Our work will also facilitate distributed deep training development with fast-adaptiveness and efficient multi-tier processing while tackling non-terrestrial system heterogeneity and dynamics.

The rest of this article is organized as follows. The following section gives the state-of-the-art. We then present serverless edge architectures and discuss ultrabroadband access and resilient networking. The final section concludes the article.

\section{The State-of-the-Art}\label{sec2} 
To integrate p-LEO or satellite swarm with ground communications, recent studies~\cite{19TVTQiu,22TWCJi,21JSACMou,21DASCXu,22WCNCRoper,22TNSMKak,Di19WC,Cao21TC} focus on spectrum sensing and sharing as well as ML-based resource allocation for such non-terrestrial coexistence. In~\cite{21DASCXu}, cognitive radio technologies are adopted to detect the channel state of primary signals and suppress co-channel interference for CubeSat swarm networks. In~\cite{22WCNCRoper}, beamspace MIMO is exploited for downline satellite swarms, requiring only position information for distributed linear precoders and a ground equalizer. Authors in~\cite{Di19WC} summarize ultra-dense LEO satellite networks and introduce satellite access architectures with supporting technologies and use cases. Furthermore, SDN-based management~\cite{19TVTQiu} provides a deep Q-learning approach to orchestrate networking, caching, and computing resources jointly for satellite-terrestrial networks. In~\cite{22TWCJi}, optimal network control structures are studied to improve the temporal control effectiveness with the least number of controllers. Authors in~\cite{21JSACMou} further consider three-dimensional (3D) terrain surface coverages by designing hierarchical unmanned aerial vehicles (UAVs) swarms via deep reinforcement learning algorithms. Moreover, an automatic network slicing platform for the Internet of space things is presented in~\cite{22TNSMKak}, which carries out various SLAs over the space-ground integrated infrastructure. Also, multi-user access schemes to non-terrestrial bases stations are investigated in~\cite{Cao21TC} via deep reinforcement learning to provide high throughput and fewer handovers for 6G traffic. Still, sensing-enabled coexistence, resilience access, and resource management developments for 6G p-LEOs are in infancy and require innovations and new approaches.

\section{Serverless Edge Architecture with Multi-Agent Deep Reinforcement Learners}\label{sec3} 

\subsection{ML-SDN Control Engines and Infrastructure} 
\begin{figure*}[!t]\centering
\includegraphics[width=6.0 in]{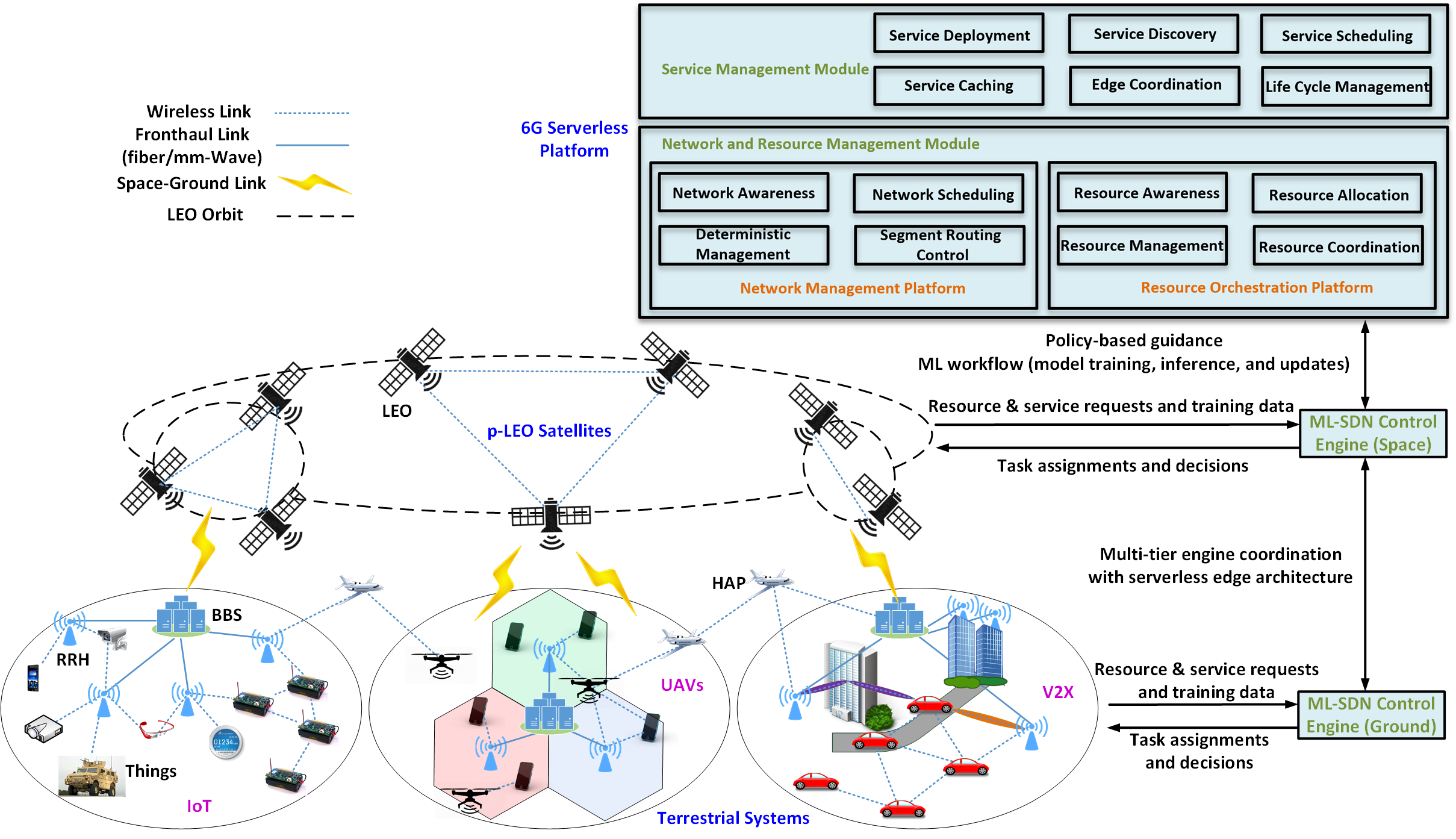}
\caption{A serverless edge architecture with multiple ML-SDN control engines for 6G p-LEO satellite networks.}\label{6gsate}
\end{figure*}

Fig.~\ref{6gsate} shows the proposed serverless edge architecture in ground and space tiers for 6G p-LEO satellite networks. The ground tier consists of several terrestrial systems, such as the Internet of things (IoT), UAVs, vehicle-to-everything (V2X), etc.; each system has a dedicated ML-SDN control engine. These control engines integrate SDN controllers with ML algorithms and manage computing, storage, and communications resources. They receive resource and service requests and training data from the serving systems and, in turn, assign tasks and control decisions back. The space tier includes p-LEO satellites from different operators, such as SpaceX, Amazon, space development agency (SDA), in orbits, and operating systems.
A serverless edge platform is established for a scalable and unified control plane to coordinate multi-tier engines and constitute a shared resource pool for virtualization and network slicing. It can alleviate the disturbance to physical infrastructure units giving time-varying resource availability and heterogeneity. Specifically, the serverless platform pushes the controllability to network edges via local control engines with edge and serverless computing. As 6G p-LEO systems cover an ultra-wide geographical and spatial area, a single centralized control with huge decision parameters is not a feasible solution. TABLE~\ref{PSU2} summarizes three control plane implementations, corresponding attributes, challenges/costs, and learning deployments for multi-tier ML-SDN engines. These multiple engines can be organized in a fully distributed, multi-domain flat, or multi-layer hierarchical manner to provide control scalability and boost learning efficiency. Notably, the multi-domain and multi-layer control can realize optimal global learning by splitting enormous optimization dimensions into collaborative engines' tasks. Such collaboration can be revamped from SDN's east-bound and west-bound application programming interfaces (APIs). The south-band APIs are expanded to help ML-SDN engines to communicate and control the respective underlying physical resources. The expanded designs are dedicated to systems' various functions, such as satellite beam steering, UAV movement control, and resource block allocation in base stations.

\begin{table*}[!t]\centering
\caption{Summary of intelligent networking infrastructures and multi-agent learning deployments.}\label{PSU2}
\begin{tabular}{|c|c|c|c|}
\hline ML-SDN infrastructures & Attribute & Challenge & Learning deployment/use cases\\
\hline Distributed control & \parbox{0.22\textwidth}{\centering Each engine regulates its domain}
                           & \parbox{0.17\textwidth}{\centering No cooperation, no centralized control}
                           & \parbox{0.19\textwidth}{\centering Individual structure; e.g.,~\cite{22TWCJi,21JSACMou,21DASCXu,22WCNCRoper,Cao21TC,lininfocomwksp21,gan2020}} \\

\hline Multi-domain control (flat)& \parbox{0.22\textwidth}{\centering Engines can cooperate with others}
                           & \parbox{0.17\textwidth}{\centering Signaling overheads among engines}
                           & \parbox{0.19\textwidth}{\centering Federated/global structures; e.g.,~\cite{19TVTQiu,22TNSMKak}} \\

\hline Multi-layer hierarchical control
                           & \parbox{0.22\textwidth}{\centering Software-defined hierarchy \& load balancing}
                           & \parbox{0.17\textwidth}{\centering More sophisticated control policies}
                           & \parbox{0.19\textwidth}{\centering Federated/hierarchical/global structures; e.g.,~\cite{21JSACMou}} \\
\hline
\end{tabular}
\end{table*}

\subsection{Serverless Edge Platform with Multi-Agent Learners} 
The proposed 6G serverless platform aims to establish an intelligent networking architecture that effectively manages computation and communications resources to meet all SLA demands in 6G p-LEO systems. As in Fig.~\ref{6gsate}, this platform adopts a Function as a Service (FaaS) and orchestrates function containers among (geographically distributed) ML-SDN control engines for highly flexible virtualization infrastructure. It provides policy-based guidances for ML workflows and consists of two crucial modules. First, the service management module enables diverse p-LEO and ML applications to request SLA portfolios (e.g., link throughput, end-to-end latency, etc.) from the lower management module. Each application is decomposed into functions (e.g., service discovery, deployment, scheduling, caching, etc.) and defines a dedicated service function chain upon the infrastructure. Second, the network and resource management module collects global network status, allocates multi-dimensional resources, and regulates multiple control engines with the ML deployments. Specifically, the network management platform maintains the topology, decides the best routes, and schedules and monitors the upper function chains. The resource orchestration platform allocates and controls underlying resources from distributed infrastructure. With the above designs, the 6G serverless platform can efficiently explore network adaptivity and realize optimal space-ground policies at serverless edges through network automation.

One primary objective of this serverless platform is to dynamically adjust ML workloads among heterogeneous computing units while considering each of their real-time capabilities. As control engines can equip with multi-agent learners, different ML deployments in TABLE~\ref{PSU2} can be implemented to ensure effective engine coordination for outstanding system performance. Specifically, a multi-agent deep reinforcement learning (MADRL) algorithm has three design components: observation processors, action predictors, and deep reinforcement learners (DRLs). An agent interacts with the environment (i.e., p-LEO networks) and obtains local observations and models (e.g., current state, action, reward, and next state). An action predictor can characterize agent collaboration by predicting other agent's actions. DRLs can derive the optimum policies based on the observation and prediction results. Furthermore, each agent/engine can also upload its local parameters to the serverless platform, training neural networks with global information. Thus, as in TABLE~\ref{PSU2}, our serverless edge platform can configure four learning deployments listed below for different coordination among ML-SDN control engines.
\begin{itemize}
\item \textbf{Individual/competitive structure}: each control engine implements its DRL from its observation processor and inactivates its action predictor by considering other engines as part of the environment. Control engines can obtain rewards and new observations stored as training experiences by interacting with the entire environment. Such historical experiences can facilitate engines to carry out ML workloads with their real-time capabilities.

\item \textbf{Federated/collaborative structure}: control engines with larger time granularity require long periods of environment observation, which degrades model training and updating performances. Collaborative deployment structures can significantly reduce such processing latency, particularly for larger control cycles. Specifically, recently emerging federated learning can be applied, where each engine trains its DRL and reports local parameters to the serverless platform. The platform then aggregates the collected parameters for a global model and disseminates the results to engines for their subsequent training. Also, this federated structure can quickly realize centralized training and decentralized execution of MADRL.

\item \textbf{Hierarchical/leader-follower structure}: Leveraging the merits of hierarchical ML-SDN infrastructures, upper/leader control engines with larger control cycles enable DRLs with their observation and predictions of lower/follower actions. Then, the followers can observe the leaders' actions and the environment; accordingly, implement their DRLs. Such hierarchical structures can also be extended with multiple layers to address scalability issues, e.g., leaders, sub-leaders, and followers.

\item \textbf{Global/joint structure}: Lastly, all control engines can upload their observations and rewards to the serverless platform for training a fully centralized DRL. Then, the platform will provide the results to engines to perform corresponding actions. In addition, historical experience storage and usage to enhance MADRL can be handled by only local engines, engines with the platform, and the sole platform,  depending on the ML-SDN infrastructures and learning deployments.
\end{itemize}

\section{Ultrabroadband Spectrum Sensing and Access} 
Due to recent advances in transceiver hardware, frequency-agile ultra-broadband reconfigurable frontend is envisioned to realize full-spectrum (1 GHz to 10 THz) sensing and communications that meet data rate, reliability, and scalability requirements. For example, the National Aeronautics and Space Administration's (NASA's) Aura satellites collect radiometric data on 118 GHz, 190 GHz, 2.5 THz, etc., in their remote sensing and earth exploration services. In 5G NR release 17, frequency ranges 1 (sub-6 GHz) and 2 (millimeter wave, mmWave) are considered to support cellular V2X (C-V2X) communications and enhance wide-area coverage. Hence, the spectrum innovation technology and multi-antenna access will need to fast, agilely, and automatically utilize the ultra-broadband to optimize physical transmissions. The network infrastructure and satellite terminals can recognize spectrum usages and exploit radio resources for communications efficiency through deep reinforcement learning based on serverless edge architectures. \emph{All-spectral analytic} for intelligent sensing and autonomous configuration thus becomes a prerequisite of optimizing spectral efficiency in 6G p-LEO networks.

\subsection{Automatic Ultrabroadband Sensing}
The main technical challenge is to develop innovative spectrum sensing techniques and sensing-informed communications for dynamic access to all-spectral resources. First, evolving from MHz to GHz spectrum characterization, wireless learning features (e.g., signal waveforms, cyclic spectrums, complex correntropy) should be extracted from raw sensory input. Accordingly, deep learning-based wideband sensing techniques (e.g., wavelet detection or compressed sensing) can be designed to identify available spectrum efficiently. Such real-time learning variants (e.g., real-time inference, fast spectrum analytic) under practical wireless channels (e.g., fast fading for highly mobile vehicles) can be further investigated for timely sensory processing. Next, these sensing algorithms can be extended with end-to-end learning techniques concerning hardware impairments, physical-layer model mismatches, and nonlinearities. An end-to-end spectrum learning with vehicular channel awareness~\cite{lininfocomwksp21} is proposed to disclose the usage behaviors of THz bands fully.
Fig.~\ref{F1} shows the performance of ultra-broadband spectrum recognition and sensing for a recent generative adversarial network (GAN)-based solution~\cite{gan2020} and our work~\cite{lininfocomwksp21}.
We set up a practical transportation environment from downtown Raleigh, North Carolina, 
and C-V2X systems with co-existing sub-6 GHz, mmWave, and THz communications.
The results imply that our scheme can effectively extract and learn multiple simultaneous connections and outperforms the GAN realization for all bands by jointly designing spectrum compression and reconstruction.

\begin{figure}\centering
\subfloat{ 
\scriptsize\begin{tabular}{|c|c|c|c|}
\hline
Downtown Raleigh area  & \multicolumn{3}{c|}{600 m$^2$ via SUMO (Simulation of Urban MObility)}                       \\ \hline
Vehicle size & \multicolumn{3}{c|}{5x1.8 m$^2$} \\ \hline
Vehicle arrival rate         & \multicolumn{3}{c|}{2.5/s}                       \\ \hline
Maximum vehicle speed            & \multicolumn{3}{c|}{55.56 m/s}                     \\ \hline
\hline
Communications band                 & sub-6GHz       & mmWave          & THz          \\ \hline
Operating frequency                & 0-2 GHz         & 26.5-29.5 GHz    & 100-550 GHz      \\ \hline
Bandwidth                         & 2 GHz             & 3 GHz             & 450 GHz          \\ \hline
Coverage                            & 100 m           & 15 m              & 15 m           \\ \hline
Antenna gain                      & 0 dBi             & 20 dBi              & 50 dBi           \\ \hline
Number of subcarriers                   & \multicolumn{3}{c|}{256}                       \\ \hline
Maximum connections          & \multicolumn{3}{c|}{8}                         \\ \hline
Guardband size                         & \multicolumn{3}{c|}{1 subcarrier}                         \\ \hline
SNR, Nyquist rate                                 & \multicolumn{3}{c|}{30 dB, 0.125}                        \\ \hline
\end{tabular}
}\\
\subfloat{
\scriptsize\begin{tabular}{|c|c|c|c|c|c|c|}
\hline              & \multicolumn{2}{c|}{sub-6 GHz}  & \multicolumn{2}{c|}{mmWave}  & \multicolumn{2}{c|}{THz}      \\ \cline{2-7}
 & \cite{gan2020} & \cite{lininfocomwksp21} & \cite{gan2020} & \cite{lininfocomwksp21} & \cite{gan2020}& \cite{lininfocomwksp21}\\ \hline
Mean square error
& 0.069   & 0.003  & 0.022   & 0.002   & 0.019  & 0.001 \\ \hline
Cosine similarity
& 0.337   & 0.995  & 0.646   & 0.996   & 0.439  & 0.991   \\ \hline
Structure similarity
& 0.303   & 0.852  & 0.739   & 0.931   & 0.629  & 0.939   \\ \hline
Detection rate
& 0\%      & 90\%    & 1.6\%   & 97\%  & 6.2\% & 94.5\%   \\ \hline
$F_1$ score
& 0\%    & 94.7\% & 3.2\% &  98.5\%  & 11.7\% & 97.1\%   \\ \hline
\end{tabular}
}
\caption{Performance comparisons of deep learning-enabled ultrabroad sensing, i.e., GAN-based~\cite{gan2020} versus our~\cite{lininfocomwksp21} schemes.}\label{F1}
\end{figure}

Moreover, deep recurrent learning, e.g., long short-term memory and gate recurrent unit, can be employed to develop dynamic all-spectral access based on time-series results from ultra-broadband sensing. Learning-based spectrum decision, sharing, and mobility can be proposed to avoid radio interference and optimize shared spectrum allocation by considering delayed sensing data. Besides, to enable radio components adaptive to time-vary environments, an autonomous frontend setting should be designed to tune channels and transmission power levels in real-time. Notably, from the designed all-spectral sensing and communications, comprehensive frontend configurations (e.g., analog electronics, bandwidth sensitivity, position) can be automatically and optimally adjusted by sensory processing and training-driven decision making (e.g., reinforcement learning).

\subsection{Fast Unsupervised MIMO Beamforming}
Two crucial challenges are supporting timely beamforming with high bandwidth for multi-antenna ultrabroad systems. First, supervised deep learning can address complicated beamformers with large antenna arrays (e.g., perfect channel state information via channel estimation). However, it is limited by the performance of labelling algorithms, which must label vast input data (especially in multi-user scenarios) during offline training. Second, hybrid beamformers need low-latency beam management to constantly provide good transmission quality in fast time-varying channels. Such channel conditions are due to peculiar LEO satellite movement and communications band (e.g., mmWave's blockage sensitivity, THz's pronounced molecular absorption, and spreading losses).

\emph{Unsupervised reinforcement learning-based beamforming} can cope with these challenges by providing fast beam tracking for the net multi-antenna gain. In serverless edge architectures, satellite swarms can coordinate for joint operations to act as a distributed antenna system.
The time-varying MIMO fading channel on space-ground communications can then be formulated by extending the $K$-user interference channel. The average downlink data rate can be obtained via medium access control (e.g., asynchronous direct-sequence code-division multiple access). Unsupervised learning requires ``differentiable" objective functions. So, we adopt a constrained user sum-rate maximization for the beamforming design. This framework will optimize power and spectrum allocations and beamformer matrices subject to maximum available power and spectrum constraints (i.e., from automatic spectrum analytic) and stochastic computation delay requirements.
The primary objective is to introduce computation-efficient unsupervised algorithms that consistently provide good beamformers adaptive to timely environmental changes. But, model-based schemes are computationally prohibited in real-time. Recent deep learning beamforming employs regularized loss functions or scaled beamforming output to satisfy power constraints, having a performance gap compared to the weighted minimum mean square error (WMMSE) counterpart. Deep unfolding techniques have been used to leverage optimization designs and residual neural networks and develop two consecutive modules for efficient unsupervised beamformers~\cite{linglobecom20}. First, a coarse estimator module is proposed to address the constrained maximization framework for outperforming WMMSE in sum-rate. Then, gradient descent beamforming is empowered with a deep unfolding module to enable fast convergence with superior performance. A balance between system resilience and rapid beam alignment/steering can also be established by upgrading unsupervised beamformers with reinforcement learning-enabled tracking. Multi-agent Q learners will be adopted with received power levels to develop new beam tracking to guarantee computation delay requirements.

\section{Robust and Resilient Satellite Networking for Service-Level Agreement Assurance}\label{sec:results} 
This section exploits the new spatial dimension of p-LEO for resilient end-to-end networking. We consider anti-jamming capabilities for space-ground communications and investigate novel space highway and cyber-harden transmissions. 

\subsection{Anti-Jammed MIMO p-LEOs}
\begin{figure}[!t]\centering
\includegraphics[width=3.0 in]{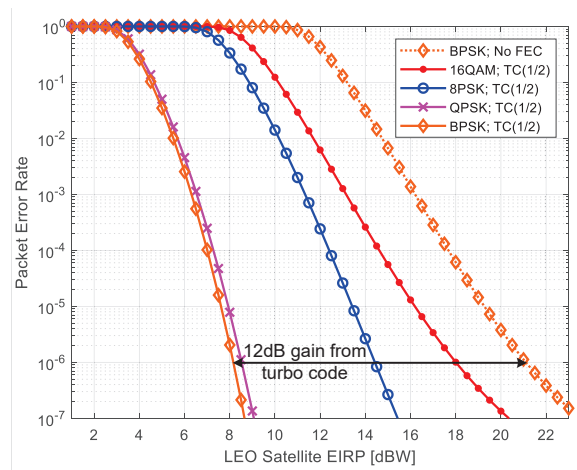}
\caption{A LEO satellite's downlink Ka-band (27.5 GHz) transmissions with different modulation and coding schemes.}\label{dlleotoground}
\end{figure}

An adaptive power control mechanism can be developed in higher bands (e.g., Ka at 26.5-40 GHz) to realize efficient ground-LEO access (particularly for uplink communications as 6G backhaul or integrated access and backhaul). This mechanism minimizes the transmit power by jointly employing satellite channel estimation and a feedback control loop. Ka-band systems (currently applied by Starlink by SpaceX) deliver substantially greater throughput than previous Ku-band offerings. Fig.~\ref{dlleotoground} shows downlink packet error rates for an LEO satellite to a line-of-sight ground terminal. Ten terminals share the channel via direct sequence spread spectrum (DSSS) with 240 (i.e., 23 dB) spreading factor, Raised Cosine pulse shape with 0.35 roll-off factor, 16 Mbps data rates, and 1,500 bits packet size. The results imply that a 12 dB gain can be obtained with turbo codes and nominal equivalent isotropically radiated power (EIRP). Also, from SDA optical communications terminals, the modulation types use on-off-keying non-return-to-zero and m-ary pulse position modulations with radio resources located at 193.1 and 195.1 THz. Hence, uplink channel estimation can be employed to enable adaptive power control schemes based on physical-layer specifications and downlink performance. For example, an auto-regressive moving-average model generally describes the dynamics of rain fading in Ka-bands. The transmit power can then vary following the fading gain to keep the signal-to-noise power ratio (SNR) level while maintaining the requested power outage probability.

Furthermore, due to the new spatial dimension in MIMO p-LEOs, such a satellite swarm can utilize line-of-sight channels and ultra-wide coverage for more robust communications with higher throughput. These MIMO LEOs adopt MIMO and spread spectrum technologies to prevent active jamming attacks or harmful co-channel interference. For example, antenna beamforming can eliminate interference by directing a null toward a jammer. In~\cite{Bosso21}, a DSSS system is revamped and tested experimentally for above 100 GHz and secure spectrum sharing, ensuring coexistence between ground THz active users and THz earth exploration satellite services.

Moreover, the anti-jamming capability of MIMO p-LEOs can be increased to the next level by exploiting their spatial dimension with MIMO beamforming and frequency hopping spread spectrum. In particular, uplink/downlink transmissions can smartly allocate specific ``virtual'' rays among total available rays through p-LEO coordination to avoid jamming signals. Larger-scale p-LEO coordination can be accomplished via hardware synchronization, timing protocol designs, or even software-defined architectures, like serverless edges. Thus, massively networked MIMO techniques can be established in satellite networks to achieve various SLAs (e.g., variable-rate MIMO links, robust and bandwidth-efficient communications, etc.).

\subsection{Cross-Constellation Space Highway} 
\begin{figure}[!t]\centering
\subfloat[Cross-constellation architecture with SBACN nodes.]{\includegraphics[width=3.4 in]{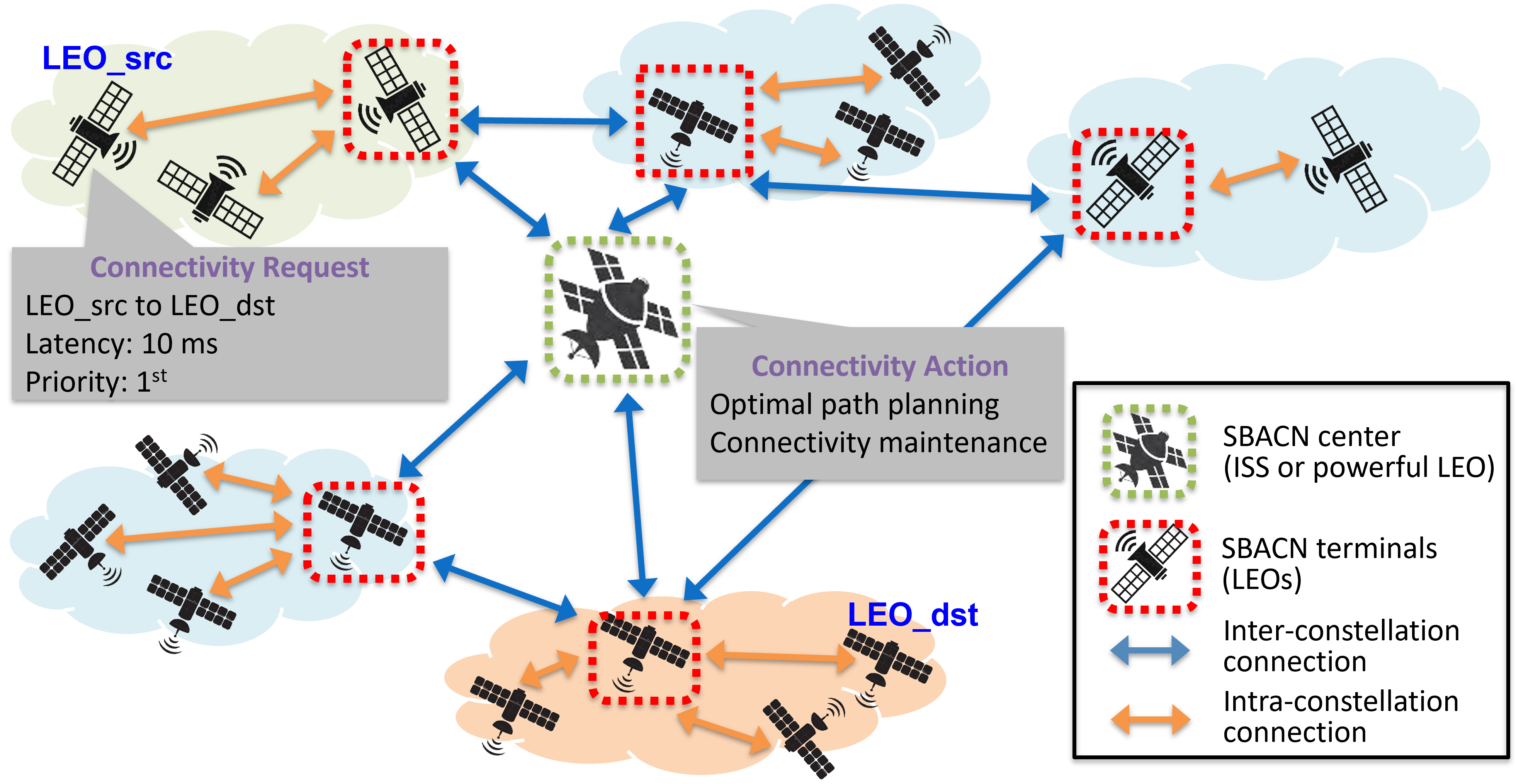}}\\
\subfloat[LEO space highway versus long multihop route.]{\includegraphics[width=3.4 in]{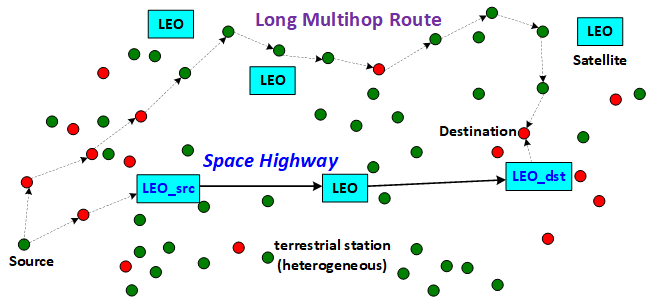}}
\caption{6G p-LEO cross-constellation space highway.}\label{hybridarch}
\end{figure}

Due to serverless edge infrastructures, heterogeneous satellite constellations from different operators, such as governments or commercial companies, can now integrate into a hybrid space architecture through space-based adaptive communications node (SBACN) and API development for cross-constellation communications command and control. As shown in Fig.~\ref{hybridarch}~(a), a hierarchical structure can be established where an SBACN center connects to several SBACN terminals. These terminals integrate existing satellite operators' constellations by acting as their gateways. SBACN terminals only need to update the resource information and service requests to the SBACN center via serverless edges and enable FaaS for connectivity and SLA satisfaction without configuring or managing resources. On the other hand, the SBACN center coordinates reliable inter-constellation-level connectivity, while each operator handles its intra-constellation-level management. The SBACN center hosts on-demand applications through dynamically instantiated containers and effectively computes optimal routing-path planning and resource allocation designs, avoiding underutilization while maintaining high throughput. Also, iteratively upgrading efforts for the APIs and underlying algorithms are negligible due to the software-defined nature.

Based on the scalable cross-constellation p-LEO platform, we can develop innovative space highway and delay-optimal multipath TCP (MPTCP) for efficient end-to-end networking.
In Fig.~\ref{hybridarch}~(b), globally scattered p-LEOs will run above terrestrial infrastructure with tremendous ground terminals. These satellites assist terrestrial communications by providing short-cut paths among ground terminals. The ground terminals now can use the upper-tier satellite network for ultra-fast data delivery rather than conventional multihop terrestrial transmissions with very-long routes. We describe research challenges for effective space highway designs as follows. Possible topologies for the p-LEOs should be investigated concerning their topological properties (e.g., network diameter, degree distribution, average hop distance) to determine the best topology with minimal latency. From our previous work with geometric random graph~\cite{Lin15m2m}, a preferable candidate could be the small-world satellite topology from the small-world phenomenon in social networks. Then, queueing network analysis and statistical quality-of-service provisioning can be applied to assure SLAs.

In addition, satellite-assisted communications suffer from long delays and frequent ground-satellite handovers (both are problematic for TCP connections). MPTCP protocols address these challenges by exploiting several Internet paths between a pair of hosts while presenting a single TCP connection to the upper layer. Thus, the upper layer applications only need to deal with a single logical master TCP connection. Multiple sub-flows are running underneath, each of which is a conventional TCP connection. To extend MPTCP to p-LEOs with serverless edge architectures, we provide a centralized delay-optimal MPTCP that minimizes the end-to-end delay with MPTCP while eliminating frequent control message signaling from distributed algorithms. Centralized controllers will facilitate (i) topology and traffic monitoring/ prediction, (ii) MPTCP awareness, and (iii) multiple disjoint paths discovery. In particular, we consider the p-LEO network topology and traffic modeling when formulating a nonlinear optimization to find end-to-end routes with minimum average delay and no congestion. We then exploit fast algorithms to provide the optimal solution within a few milliseconds, thus achieving timely and reliable p-LEO satellite systems.

\subsection{Cyber-Harden Opportunistic Multipath Routing} 
For connectivity robustness and resilience, defensive and self-healing algorithms should be designed to tackle possible cyber-attacks and provide ceaseless connections in node failures, deep fades, and malicious behaviors. A straightforward solution is to create backup paths, but it is resource prohibitive in assigning wireless backup capacity. Hence, we are revamping our opportunistic multipath routing algorithms~\cite{Lin16TMC,Lin16IoT} to provide cyber-harden p-LEO communications. In~\cite{Lin16TMC}, a cognitive and opportunistic relay solution is designed for reliable communications and connections in a machine swarm. This design enables machines to cognize and adapt to environments for mitigating inter-system interference with existing networks and realizes opportunistic selections of cooperative relay machines based on link qualities. We extend with p-LEOs' peculiarities to distributedly concatenate link transmissions for satellite swarm resilience.
In~\cite{Lin16IoT}, ``virtual'' MIMO at the session level is established, and probabilistic network-coded routing is developed for large-scale cognitive machine swarms. This work enables spatial multiplexing and diversity with session-level traffic and ensures end-to-end delay by employing network coding techniques with underlaid routing algorithms.
The developed solution expands multi-user MIMO into multihop multipath transmissions. 
We thus exploit such dynamic cooperation for the fault tolerance of p-LEO systems, robust to node or link failures.

\begin{remark}[Practical Experimentation]
In February 2022, Lockheed Martin has awarded an SDA's Tranche 1 Transport Layer contract to demonstrate an interoperable, connected, secure mesh network of 42 LEO satellites that link terrestrial warfighting domains to space sensors. As deeply involved in this development, we conduct the proof-of-concept of this paper's solutions for anti-jammed p-LEO satellites and cross-constellation communications. Our designs' on-orbit deployment is expected to be in 2024-2026. We also expand Cisco edge AI with our serverless edge framework in 
telemedicine and broadband connectivity in rural areas.
\end{remark}

\section{Conclusions}\label{sec:conclusion}
LEO mega-constellations have promised to serve isolated or remote communities and fulfill the needs of landlocked areas with limited infrastructure investments. However, there is still little work that simultaneously addresses the satellite access and inter-tier networking within the 6G context and provides a coherent serverless architecture for such ecosystems. This article introduces unique serverless edge architectures with multi-tier deep reinforcement learners and emphasizes necessary architectural, management, and operational advances, thus bringing a new frontier for resilient access equality.

\bibliographystyle{IEEEtran}
\bibliography{ref}

\begin{thebibliography}{10}
\providecommand{\url}[1]{#1}
\csname url@samestyle\endcsname
\providecommand{\newblock}{\relax}
\providecommand{\bibinfo}[2]{#2}
\providecommand{\BIBentrySTDinterwordspacing}{\spaceskip=0pt\relax}
\providecommand{\BIBentryALTinterwordstretchfactor}{4}
\providecommand{\BIBentryALTinterwordspacing}{\spaceskip=\fontdimen2\font plus
\BIBentryALTinterwordstretchfactor\fontdimen3\font minus
  \fontdimen4\font\relax}
\providecommand{\BIBforeignlanguage}[2]{{%
\expandafter\ifx\csname l@#1\endcsname\relax
\typeout{** WARNING: IEEEtran.bst: No hyphenation pattern has been}%
\typeout{** loaded for the language `#1'. Using the pattern for}%
\typeout{** the default language instead.}%
\else
\language=\csname l@#1\endcsname
\fi
#2}}
\providecommand{\BIBdecl}{\relax}
\BIBdecl

\bibitem{22TWCJi}
S.~Ji, D.~Zhou, M.~Sheng, and J.~Li, ``Mega satellite constellation system
  optimization: From a network control structure perspective,'' \emph{IEEE
  Trans. Wireless Commun.}, vol.~21, no.~2, pp. 913--927, 2022.

\bibitem{21JSACMou}
Z.~Mou, Y.~Zhang, F.~Gao, H.~Wang, T.~Zhang, and Z.~Han, ``Deep reinforcement
  learning based three-dimensional area coverage with uav swarm,'' \emph{{IEEE}
  J. Sel. Areas Commun.}, vol.~39, no.~10, pp. 3160--3176, 2021.

\bibitem{21DASCXu}
C.~Xu, T.~Yang, and H.~Song, ``Spectrum sensing of cognitive radio for cubesat
  swarm network,'' in \emph{IEEE/AIAA DASC}, 2021, pp. 1--8.

\bibitem{22WCNCRoper}
M.~R\"{o}per, B.~Matthiesen, D.~W\"{u}bben, P.~Popovski, and A.~Dekorsy,
  ``Beamspace mimo for satellite swarms,'' in \emph{IEEE WCNC}, Apr 2022.

\bibitem{22TNSMKak}
A.~Kak and I.~F. Akyildiz, ``Towards automatic network slicing for the internet
  of space things,'' \emph{IEEE Trans. Netw. Service Manag}, vol.~19, no.~1,
  pp. 392--412, 2022.

\bibitem{Di19WC}
B.~Di, L.~Song, Y.~Li, and H.~V. Poor, ``Ultra-dense leo: Integration of
  satellite access networks into 5g and beyond,'' \emph{{IEEE} Trans. Wireless
  Commun.}, vol.~26, no.~2, pp. 62--69, 2019.

\bibitem{19TVTQiu}
C.~Qiu, H.~Yao, F.~R. Yu, F.~Xu, and C.~Zhao, ``Deep q-learning aided
  networking, caching, and computing resources allocation in software-defined
  satellite-terrestrial networks,'' \emph{{IEEE} Trans. Veh. Technol.},
  vol.~68, no.~6, pp. 5871--5883, 2019.

\bibitem{Cao21TC}
Y.~Cao, S.-Y. Lien, and Y.-C. Liang, ``Deep reinforcement learning for
  multi-user access control in non-terrestrial networks,'' \emph{{IEEE} Trans.
  Commun.}, vol.~69, no.~3, pp. 1605--1619, 2021.

\bibitem{lininfocomwksp21}
C.-H. Lin, S.-C. Lin, and E.~Blasch, ``Tulvcan: Terahertz ultra-broadband
  learning vehicular channel-aware networking,'' in \emph{IEEE INFOCOM WKSHPS},
  2021, pp. 1--6.

\bibitem{gan2020}
X.~Meng, H.~Inaltekin, and B.~Krongold, ``End-to-end deep learning-based
  compressive spectrum sensing in cognitive radio networks,'' in \emph{IEEE
  ICC}, 2020, pp. 1--6.

\bibitem{linglobecom20}
C.-H. Lin, Y.-T. Lee, W.-H. Chung, S.-C. Lin, and T.-S. Lee, ``Unsupervised
  resnet-inspired beamforming design using deep unfolding technique,'' in
  \emph{IEEE GLOBECOM}, 2020, pp. 1--7.

\bibitem{Bosso21}
C.~Bosso, P.~Sen, X.~Cantos-Roman, C.~Parisi, N.~Thawdar, and J.~M. Jornet,
  ``Ultrabroadband spread spectrum techniques for secure dynamic spectrum
  sharing above 100 ghz between active and passive users,'' in \emph{IEEE
  DySPAN}, 2021, pp. 45--52.

\bibitem{Lin15m2m}
S.-C. Lin, L.~Gu, and K.-C. Chen, ``Statistical dissemination control in large
  machine-to-machine communication networks,'' \emph{IEEE Trans. Wireless
  Commun.}, vol.~14, no.~4, pp. 1897--1910, 2015.

\bibitem{Lin16TMC}
S.-C. Lin and K.-C. Chen, ``Cognitive and opportunistic relay for qos
  guarantees in machine-to-machine communications,'' \emph{{IEEE} Trans. Mobile
  Comput.}, vol.~15, no.~3, pp. 599--609, 2016.

\bibitem{Lin16IoT}
------, ``Statistical qos control of network coded multipath routing in large
  cognitive machine-to-machine networks,'' \emph{IEEE Internet Things J.},
  vol.~3, no.~4, pp. 619--627, 2016.

\end{thebibliography}

\begin{IEEEbiography}
{Shih-Chun~Lin} [M'17] (slin23@ncsu.edu) received his Ph.D. degree from the Georgia Institute of Technology in 2017. He is currently an assistant professor with the Department of Electrical and Computer Engineering, North Carolina State University, leading the Intelligent Wireless Networking Laboratory. His research interests include 6G networks, software-defined infrastructure, machine learning techniques, mathematical optimization, and performance evaluation.
\end{IEEEbiography}

\begin{IEEEbiography}
{Chia-Hung Lin} (clin25@ncsu.edu) received a B.S. degree in Electrical Engineering from Chang Gung University, Taoyuan, Taiwan, in 2016, and an M.S. degree at the Institute of Communications Engineering from National Sun Yat-Sen Kaohsiung, Taiwan, in 2018. He is currently a Ph.D. student at the Department of Electrical and Computer Engineering, North Carolina State University.
\end{IEEEbiography}

\begin{IEEEbiography}
{Liang C. Chu} (liang.c.chu@lmco.com) is a Lockheed Martin Fellow at Lockheed Martin Space Systems Company (LMSSC), Lockheed Martin Corporation. He has been a Principal Investigator for many governmental and Lockheed Martin IRAD programs. He has led numerous LMSSC satellite communication systems development. Before joining LMSSC, he worked in IBM and Bell South/Science and Technology Organization. He has over 30 years of experience in communications, digital signal processing, and wireless networking. 
He received his Ph.D. in Electrical and Computer Engineering from the Georgia Institute of Technology.
\end{IEEEbiography}

\begin{IEEEbiography}
{Shao-Yu Lien} (sylien@ccu.edu.tw) received his B.S. degree from National Taiwan Ocean University in 2004, his M.S. degree from National Cheng Kung University in 2006, and his Ph.D. degree from National Taiwan University in 2011. He has been with the Department of Computer Science and Information Engineering, National Chung Cheng University, Taiwan, since 2017. His current research interests include 5G/6G mobile networks, cyber-physical systems, and configurable networks. He received the IEEE ComSoc Asia-Pacific Outstanding Paper Award 2014, the Scopus Young Researcher Award 2014, and the URSI AP-RASC 2013 Young Scientist Award. 
\end{IEEEbiography}

\end{document}